\documentclass[twocolumn,showpacs,amsmath,amssymb]{revtex4-1}

\usepackage{graphicx,amsmath,amssymb,bm}
\usepackage{amsthm}
\usepackage{amsfonts}
\usepackage{dcolumn}
\usepackage{bm}
\usepackage{mathrsfs}
\usepackage[usenames,dvipsnames]{color}
\usepackage{amsfonts,mathrsfs,amssymb}
\usepackage{braket}

\usepackage[colorlinks,linkcolor=blue,anchorcolor=blue,citecolor=blue]{hyperref}

\newcommand{\bq}{{\mathbf{q}}}

\usepackage{shortcuts}

\newcommand{\nord}[1]{\ensuremath\left\{#1\right\}}

\newcommand{\red}[1]{\textcolor{red}{#1}}

\begin{document}

\title{Generator-coordinate reference states for spectra and $0\nu\beta\beta$
decay in the in-medium similarity renormalization group}

\author{J. M. Yao}
  \affiliation{Department of Physics and Astronomy, University of North Carolina, Chapel Hill, North Carolina 27516-3255, USA}
\affiliation{FRIB/NSCL, Michigan State University, East Lansing, Michigan 48824,
USA}

\author{J. Engel}
  \affiliation{Department of Physics and Astronomy, University of North Carolina, Chapel Hill, North Carolina 27516-3255, USA}
\author{C. F. Jiao}
  \affiliation{Department of Physics and Astronomy, University of North Carolina, Chapel Hill, North Carolina 27516-3255, USA}
  \affiliation{Department of Physics, San Diego State University, San Diego,
  Caniformia 92182-1233, USA} \date{\today}

\author{L. J. Wang}
  \affiliation{Department of Physics and Astronomy, University of North
  Carolina, Chapel Hill, North Carolina 27516-3255, USA}

\author{H. Hergert}
\affiliation{FRIB/NSCL and Department of Physics and Astronomy, Michigan State
University, East Lansing, Michigan 48824, USA}

\date{\today}

 \begin{abstract}

We use a reference state based on symmetry-restored states from deformed
mean-field or generator-coordinate-method (GCM) calculations in conjunction with
the in-medium similarity-renormalization group (IMSRG) to compute spectra and
matrix elements for neutrinoless double-beta ($0\nu\beta\beta$) decay.  Because
the decay involves ground states from two nuclei, we use evolved operators from
the IMSRG in one nucleus in a subsequent GCM calculation in the other.  We
benchmark the resulting IMSRG+GCM method against complete shell-model diagonalization 
for both the energies of low-lying states in $^{48}$Ca and $^{48}$Ti and the 
$0\nu\beta\beta$ matrix element for the decay of $^{48}$Ca, all in a single 
valence shell.  Our approach produces better spectra than either the IMSRG with a
spherical-mean-field reference or GCM calculations with unevolved operators.
For the $0\nu\beta\beta$ matrix element the improvement is slight.


 \end{abstract}
 
\pacs{21.60.Jz, 24.10.Jv, 23.40.Bw, 23.40.Hc}
\maketitle


\section{Introduction} 

The search for neutrinoless double-beta ($0\nu\beta\beta$) decay is an important
effort in modern nuclear and particle physics, in part because it offers the only
real hope of determining whether neutrinos are Majorana
particles~\cite{AEE08,Vogel12}.  The rate of decay, however, depends on nuclear
matrix elements that must be accurately calculated to allow experimentalists to
plan efficiently and interpret results. At present, the matrix elements
predicted by various nuclear models
\cite{Menendez09,Rodriguez10,Vaquero13,Mustonen13,Simkovic13,Hinohara14,
Hyvarinen15, Barea15, Yao15, Iwata16, Senkov16,Jiao17,Fang18} differ by factors of
up to 3~\cite{EM17}.  Theorists have worked hard to identify the shortcomings of
the models and improve them accordingly.  Ultimately, however, we will need
fully ab-initio calculations with controlled uncertainty.

Ab-initio methods have improved rapidly in recent years
\cite{Lee09,Navratil09,Soma13,Barrett13,Hagen14,Carlson15,Draayer16,Hergert16}.  Most
applications, however, are still in relatively light nuclei near closed shells.
The nuclei used in $\beta\beta$ experiments, among them $^{76}$Ge, $^{82}$Se, $^{130}$Te,
$^{100}$Mo, $^{136}$Xe, and $^{150}$Nd, are typically heavier and often far from
closed shells in protons, neutrons, or both.  Among the existing ab-initio
methods, the in-medium similarity renormalization group (IMSRG) method
\cite{Tsukiyama11,Hergert16,Hergert17} is particularly suited to an extension to
such mid-shell nuclei.  One scheme for making the extension involves choosing
the generators of the RG flow to decouple a shell-model space from the rest of
the full many-body Hilbert space
\cite{Tsukiyama12,Bogner14,Stroberg16,Stroberg17}. Although the framework,
called the valence-space IMSRG, has been used to describe nuclei as heavy as tin
\cite{Morris18}, it suffers from the use of a closed-shell reference state or a
spherical reference ensemble, both of which omit collective correlations
\cite{Parzuchowski17}. Such correlations are difficult to capture in an
approximate SRG flow that simplifies induced many-body operators. 

In the IMSRG as currently practiced, induced $A$-body operators with $A>2$ are
included only approximately by retaining just their normal-ordered one- and
two-body pieces.  Collective effects will be better represented if they are
explicitly built into the reference state.  To use a more general reference
state, one must extend the procedure of normal ordering.  Refs.\
\cite{Kutzelnigg97,Mukherjee97,Kong10} show how to define a normal ordering that
applies to any reference state; the work of Refs.\ \cite{Hergert13,Hergert14}
made use of the scheme with a number-projected spherical Hatree-Fock-Bogoliubov
reference state (which explicitly includes pairing correlations) to apply the
IMSRG to spherical open-shell isotopes.  More recently, the authors of Ref.\
\cite{Gebrerufael17} used a no-core shell-model reference state in just a few
shells (for lighter nuclei).  They showed that the IMSRG flow with respect to
that reference generates a Hamiltonian for subsequent calculations in the same
few shells that effectively incorporates the physics from many higher shells.

In this paper, we generalize the reference state even further, not only by
including angular-momentum-projected Hartree-Fock-Bogoliubov (HFB) states with
deformation, but also by using the generator-coordinate method (GCM) to mix many
such states, so that the IMSRG can be applied to essentially any nucleus.  The
GCM is flexible enough to include in the reference both the ``static''
correlations associated with collectivity --- superfluidity and deformation ---
and ``dynamic'' correlations associated with shape fluctuations.  The IMSRG flow
then incorporates non-collective correlations, generating an improved
Hamiltonian that we use in a second GCM calculation and evolved transition
operators with which to obtain other nuclear properties.  Here we focus not only
on energy spectra, but also on the
$0\nu\beta\beta$ transition matrix elements, and show how to include the effects
of complicated non-collective correlations in that process.  We then benchmark
the method against the conventional shell model for the spectra of $^{48}$Ca and
$^{48}$Ti and the transition matrix element between the two.
 
The paper is organized as follows.  In Sec.\ \ref{Formalism} we present the
IMSRG+GCM method for computing both the energies of low-lying states and the
matrix elements for $0\nu\beta\beta$ decay, and use $^{48}$Ca and $^{48}$Ti
within a valence shell to illustrate the method. Section \ref{Results} presents
and discusses the results.  Section \ref{Summary} offers a summary and some
perspective.

\section{Formalism}

\label{Formalism}
In this section we present a general framework for the IMSRG+GCM. Although we
restrict our calculations to a single shell with a phenomenological Hamiltonian
here, all the expressions we develop are more general.  We will report their
application within an ab-initio calculation, with interactions from chiral
effective field theory, in a separate paper.

\subsection{The IMSRG}

The basic idea of the IMSRG is to use a flow equation to gradually decouple a
chosen reference state $\ket{\Phi}$ (or more generally a space or ensemble) from
all other states.  One defines a Hamiltonian $H(s)$ that depends on a flow
parameter $s$ as
\beq
\label{Hs}
 \hat H(s)=\hat U(s)\hat H_0\hat U^\dagger(s) \,,
\eeq
with $\hat U(0)=1$, where $\hat H_0$ is the initial Hamiltonian and $\hat U(s)$
represents a set of continuous unitary transformations that drive $\hat H_0$ to
a specific form, e.g., by eliminating certain matrix elements or minimizing its
expectation value.  Taking the derivative $d/ds$ of both sides of Eq.\
\eqref{Hs} yields the flow equation 
\beqn
\label{flow-H}
\dfrac{d\hat H(s)}{ds} = [\hat \eta(s), \hat H(s)] \,,
\eeqn
where we have introduced the anti-Hermitian generator of the transformation,
\beq
\hat \eta(s)\equiv\dfrac{d\hat U(s)}{ds}\hat U^\dagger (s) \,.
\eeq

Supposing the Hamiltonian $\hat H$ --- either $\hat H_0$ or an approximate $\hat
H(s)$ --- is composed of one-body, two-body and three-body terms, and writing
strings of creation and annihilation operators as
\beq
  A^{pqr\ldots}_{stu\ldots} = a^\dagger_pa^\dagger_qa^\dagger_r\ldots a_u a_t a_s\,,
\eeq
we have
\beq
\hat H = \sum_{pq} t^p_{q} A^p_q  + \dfrac{1}{4}\sum_{pqrs} V^{pq}_{rs}
A^{pq}_{rs}
+\dfrac{1}{36}\sum_{pqrstu}W^{pqr}_{stu} A^{pqr}_{stu} \,.
\eeq
Using the generalized normal ordering of Kutzelnigg and Mukherjee
\cite{Kutzelnigg97,Mukherjee97,Kong10}, we can normal-order $\hat H$ with respect to
our arbitrarily chosen reference state $\ket{\Phi}$:
\beqn
\label{normal-ordered-H}
\hat H &=& E+ \sum_{pq} f^{p}_{q} \nord{A^p_q}
+ \dfrac{1}{4}\sum_{pqrs}  \Gamma^{pq}_{rs} \nord{A^{pq}_{rs}}\nonumber\\
&& + \dfrac{1}{36}\sum_{pqrstu} W^{pqr}_{stu}  \nord{A^{pqr}_{stu}} \,.
\eeqn
By definition, the expectation values of normal-ordered operators, indicated by
$\nord{A^{p\ldots}_{q\ldots}}$, with respect to the reference state are zero.
Thus, the normal-ordered zero-body term corresponds to the reference-state
energy $E$, which is given by
\beqn
\label{H:0b}
 E=\bra{ \Phi} \hat H\ket{ \Phi}
&=&\sum_{pq} t^p_{q}\rho^p_q
+\dfrac{1}{4}\sum_{pqrs}  V^{pq}_{rs} \rho^{pq}_{rs}\nonumber\\
&& +\dfrac{1}{36}\sum_{pqrstu}W^{pqr}_{stu}\rho^{pqr}_{stu} \,.
\eeqn
The normal-ordered one-body and two-body terms are
\beqn
\label{H:1b}
 f^{p}_{q}&=& t^{p}_{q}  +  \sum_{rs} V^{pr}_{qs} \rho^r_s
+\dfrac{1}{4}\sum_{rstu} W^{prs}_{qtu}\rho^{rs}_{tu} \,,\\
\label{H:2b}
 \Gamma^{pq}_{rs}  &=& V^{pq}_{rs}+ \sum_{tu} W^{pqt}_{rstu}\rho^{t}_{u}\,.
\eeqn
In Eqs.\ \eqref{H:0b}--\eqref{H:2b}, we have introduced the usual density
matrices 
\bsub
\beqn
\rho^p_q &=& \bra{ \Phi} A^{p}_{q}\ket{\Phi}\,,\\
\rho^{pq}_{rs} &=& \bra{ \Phi} A^{pq}_{rs}\ket{\Phi}\,,\\
\rho^{pqr}_{stu} &=& \bra{ \Phi} A^{pqr}_{stu}\ket{\Phi}\,.
\eeqn
\esub 
Correlations within the reference state are encoded in the corresponding
\emph{irreducible} density matrices (also referred to as cumulants): 
\bsub
\label{cumulants}
\beqn
\lambda^p_q &=&  \rho^p_q\,, \\
\lambda^{pq}_{rs} &=& \rho^{pq}_{rs}  - {\cal A}(\lambda^p_r\lambda^q_s)
 = \rho^{pq}_{rs}  - \lambda^p_r\lambda^q_s +  \lambda^p_s\lambda^q_r\,,\\
\lambda^{pqr}_{stu} &=& \rho^{pqr}_{stu} - {\cal
A}(\lambda^p_s\lambda^{qr}_{tu}+\lambda^p_s\lambda^{q}_{t}\lambda^{r}_{u}) \,,
\eeqn
\esub
where the antisymmetrization operator ${\cal A}$ generates all possible
permutations (each only once) of upper indices and lower indices. For
independent particle states, the two-body irreducible density vanishes and we
recover the usual factorization of many-body density matrices into
antisymmetrized products of the one-body density matrix.

To decouple $\ket{\Phi}$, one usually chooses an appropriate generator $\hat
\eta$ and then solves a set of coupled ordinary differential equations (ODEs),
derived from Eq.\ \eqref{flow-H}, for $\hat f,\hat \Gamma,\ldots$
\cite{Tsukiyama11,Hergert13}. Instead, however, one can solve a similar flow
equation for the unitary transformation operator $\hat U(s)$,
\beq\label{eq:flow_U}
\dfrac{d\hat U(s)}{ds} = \hat \eta(s) \hat U(s)\,,
\eeq
whose solution can formally be written in terms of the ${\cal S}$-ordered
exponential
\beq
\hat U(s) = {\cal S}\exp \int^s_0 ds' \hat \eta(s') \,,
\eeq
which is short-hand for the Dyson series expansion of $\hat U(s)$.  As shown
first by Magnus \cite{Magnus1954,Blanes2009}, if certain convergence
conditions are satisfied it is possible to write $\hat U(s)$ as a proper
exponential of an anti-Hermitian operator $\hat \Omega(s)$:
\beq
\hat U(s)\equiv e^{\hat \Omega(s)}\,.
\eeq
Equation (\ref{eq:flow_U}) can then be re-expressed as a flow equation for $\hat
\Omega$: 
\beq
\label{flow_omega}
\dfrac{d\hat \Omega(s)}{ds}=\sum^\infty_{n=0} \dfrac{B_n}{n!} [\hat \Omega(s),\hat \eta(s)]^{(n)} \,, 
\eeq
where we define nested commutators as
\bsub\beqn
  \left[\hat \Omega(s),\hat \eta(s)\right]^{(0)} &=& \hat \eta(s)\,,\\ 
  \left[\hat \Omega(s),\hat \eta(s)\right]^{(n)} &=& \left[\hat \Omega(s), \left[\hat \Omega(s),\hat \eta(s)\right]^{(n-1)}\right]\,,
\eeqn\esub
and $B_{n=0,1,2,3,\cdots}$ are the Bernoulli numbers $\{ 1, -1/2, 1/6, 0, \cdots
\}$.  As discussed in Ref.~\cite{Morris15}, the reformulation of the IMSRG via
the Magnus expansion has two major advantages. First, the anti-Hermiticity of
$\hat \Omega$ guarantees that $\hat U(s)$ is unitary throughout the flow, even
when low-order numerical ODE solvers are used to integrate
Eq.~\eqref{flow_omega}. Second, it greatly facilitates the evaluation of
observables. In the traditional approach, we would need to solve flow equations
for each additional operator \emph{simultaneously} with Eq.~\eqref{flow-H}
because of the dynamical nature of the generator, while $\hat \Omega(s)$ allows
us to construct arbitrary evolved operators by using the
Baker-Campbell-Hausdorff (BCH) formula:
\beq
\label{BCH}
\hat O(s) = e^{\hat \Omega(s)} \hat O(0)  e^{-\hat \Omega(s)} = 
  \sum^\infty_{n=0} \dfrac{1}{n!} [\hat \Omega(s),\hat O(0)]^{(n)}
\,.
\eeq

As mentioned earlier, the IMSRG generator $\hat \eta(s)$ is chosen to implement
a specific decoupling. For closed-shell nuclei, the ability to use an
uncorrelated reference allows us to distinguish particle and hole states, which
simplifies the formulation of decoupling conditions \cite{Hergert16,Hergert17},
and the subsequent construction of $\hat \eta(s)$. For correlated reference
states like those we aim to use here, this distinction is lost, and one needs to
carefully consider the proper generalization of the generator. Here, we use the
Brillouin generator, which is essentially the gradient of the energy under a
general unitary transformation (see Appendix~\ref{append1} and
Ref.~\cite{Hergert17}):
\bsub
\beqn
\eta^p_q &\equiv& \bra{\Phi} \left[ \hat H, \nord{A^p_q} \right] \ket{\Phi}\,,\\
\eta^{pq}_{rs} &\equiv& \bra{\Phi} \left[ \hat H, \nord{A^{pq}_{rs}} \right] \ket{\Phi}\,.
\eeqn
\esub
To implement the IMSRG flow either in the traditional (Eq.~\eqref{flow-H})
or Magnus formulations (Eq.~\eqref{flow_omega}), we need to close the system
of flow equations by truncating the operators at a given particle rank.  We
adopt the IMSRG(2) approximation and truncate $\hat H(s), \hat \eta(s)$, and
$\hat  \Omega(s)$, as well as all commutators, at the normal-ordered two-body
level. This is consistent with the so-called NO2B approximation that is applied
to the input Hamiltonian in a variety of many-body approaches (see, e.g.,
\cite{Hagen2007,Roth2012,Gebrerufael2016,Hagen2016}).  With this choice of
operator truncation, up to three-body irreducible density matrices of the
reference states appear in the Brillouin generator and the flow equations. We
will show that the irreducible three-body density in the Brillouin generator is
vital to the convergence of the IMSRG(2) flow equations.

\subsection{\label{sec:ref}Choice of reference state}

We would like to explore reference states $\ket{\Phi}$ that incorporate
collective (or ``static'') correlations, such as those associated with pairing
and deformation, plus fluctuations in some of these collective quantities.  To
include such correlations, we use the GCM to find an optimal linear combination
of deformed HFB states (distinguished from one another by a set of coordinates
$\bq$), projected onto states with both well-defined neutron ($N$) and proton
($Z$) number and angular momentum $J=0$:
\beq
\ket{\Phi^{J=0}_{\alpha}} = \sum_{\bq} f^{J=0}_\alpha (\bq) \ket{NZJ=0, \bq}\,,
\eeq
where $\alpha$ denotes a particular linear combination and the non-orthogonal
basis states in which the GCM states are expanded are given by
\beq
\ket{NZJ=0 (\bq)} = \hat P^N  \hat P^Z \hat P^{J=0}_{00} \ket{\bq} \,.
\eeq
Here, the particle-number projection operator is 
\beq
\hat P^{\tau} = \dfrac{1}{2\pi}\int^{2\pi}_0 d\varphi_{\tau}  e^{i(\hat
N_\tau-N_\tau)\varphi_{\tau}} \,, 
\eeq
with $\hat N_\tau$ the particle-number operator for either neutrons ($\tau=n$)
or protons ($\tau=p$), and the angular-momentum projection operator is
\beq
\hat P^{J}_{MK}=\dfrac{2J+1}{8\pi^2}\int d\Omega D^{J\ast}_{MK}(\Omega) \hat
R(\Omega) \,,
\eeq
with $D^{J}_{MK}(\Omega)$ a Wigner-D function. The projector $\hat P^J_{MK}$ 
extracts from the intrinsic state $\ket{\bq}$ the component whose
angular momentum along the intrinsic $z$ axis is given by $K$.  In the following, 
we restrict ourselves to axially-symmetric deformation, and thus $K=0$. 

We obtain the weight function $f^{J}_\alpha(\bq)$ from the variational
principle, which leads to the Hill-Wheeler-Griffin equation~\cite{Ring80}:
\begin{equation}\label{HWG}
\sum_{\bq_b} \left[  \mathscr{H}^{J}_{\bq_a, \bq_b}-E^{J}_{\alpha}
\mathscr{N}^{J}_{\bq_a, \bq_b}\right]f^{J}_\alpha(\bq_b) =0 \,.
\end{equation}
The Hamiltonian kernel $ \mathscr{H}^{J}_{\bq, \bq_b}$ and norm kernel
$\mathscr{N}^{J}_{\bq_a, \bq_b}$ are given by
\begin{equation}
\label{kernel}
 \mathscr{O}^{J}_{\bq_a, \bq_b}
 =\bra{ NZ J (\mathbf{q}_a)}  \hat O \ket{ NZ J  (\mathbf{q}_b)} \,,
\end{equation}
with the operator $\hat O$ representing either $\hat H$ or $1$.

\subsection{Matrix elements for the $0\nu\beta\beta$ decay}

Let us now consider the evaluation of the matrix element for the
$0\nu\beta\beta$ decay of an initial nuclear state $\ket{ \Psi_I(0^+_1)}$ to a
final state $\ket{ \Psi_F(0^+_1)}$, 
\beqn
M^{0\nu} = \bra{ \Psi_F(0^+_1)} \hat O^{0\nu}(0) \ket{ \Psi_I(0^+_1)}.
\eeqn
Here, $\hat O^{0\nu}(0)$ is the bare, unevolved two-body transition operator
\cite{Rodriguez10,Hinohara14,Yao15,Jiao17} whose form is given by
\beq\label{eq:trans_op}
\hat O^{0\nu}(0)
=\red{\dfrac{1}{4}} \sum_{pp'nn'} O^{pp'}_{nn'}  \nord{A^{pp'}_{nn'}} \,,
\eeq
where $p, p'$ and $n, n'$ are indices for proton and neutron states,
respectively.  

In the IMSRG+GCM approach, we represent the initial and final states as
$\ket{\Psi_{I/F}}=e^{-\hat\Omega_{I/F}(s)}\ket{\Phi_{I/F}}$, where the unitary
transformations capture correlations that are missing from the GCM wave
functions $\ket{\Phi_{I/F}}$. One can readily show that the GCM wave functions
are solutions to the Schr\"odinger equations for the evolved Hamiltonian
operators,
\beq\label{eq:effective_schroedinger}
\hat H_{I/F}(s) \ket{\Phi_{I/F}} = E \ket{\Phi_{I/F}}\,,
\eeq
up to IMSRG truncation errors (cf.~Refs.~\cite{Hergert16,Hergert17}).

The transition matrix element now reads
\beq\label{eq:me}
M^{0\nu}(s) = \bra{ \Phi_F} e^{\hat \Omega_F(s)} \hat O^{0\nu}(0) e^{-\hat \Omega_I(s)}
\ket{ \Phi_I} \,,
\eeq
and we encounter two complications. The first is that $\hat \Omega_F(s)$ and
$\hat \Omega_I(s)$ are normal-ordered with respect to different reference
states; this difficulty can be overcome by re-normal ordering all operators with
respect to a common reference. The second, more challenging complication is that
the difference between $\hat \Omega_I(s)$ and $\hat \Omega_F(s)$ prevents us
from using a straightforward BCH expansion to evaluate the matrix element. 
To proceed, we note that we can rewrite Eq.~\eqref{eq:me} either as
\beqn\label{eq:me_pf_a}
M^{0\nu}(s) &=& \bra{ \Phi_F}e^{\hat \Omega_F(s)}e^{-\hat \Omega_I(s)}  e^{\hat
\Omega_I(s)} \hat O^{0\nu}(0) e^{-\hat \Omega_I(s)} \ket{\Phi_I}\nonumber\\
 &=& \bra{ \Phi_F}e^{\hat \Omega_F(s)}e^{-\hat \Omega_I(s)}  \hat O_I^{0\nu}(s) \ket{\Phi_I}
\eeqn
or
\beq\label{eq:me_pi_a}
M^{0\nu}(s) = \bra{ \Phi_F}\hat O_F^{0\nu}(s) e^{\hat \Omega_F(s)}e^{-\hat
\Omega_I(s)} \ket{\Phi_I}
\eeq
with $ \hat O_{I/F}^{0\nu}(s) =e^{\hat \Omega_{I/F} } \hat O^{0\nu}e^{-\hat
\Omega_{I/F} }$.  Inspecting the unitary transformations acting on the initial
GCM wave function in the previous equation, we define
\beq\label{eq:bar_state_pi}
\ket{\overline\Phi_I} \equiv e^{\hat \Omega_F(s)}e^{-\hat \Omega_I(s)}
\ket{\Phi_I} = e^{\hat \Omega_F(s)} \ket{\Psi_I}\,,
\eeq
so that we have the unitary transformation for the final nucleus acting on an
eigenstate of the initial nucleus. An analogous definition for the final nucleus
results from Eq.~\eqref{eq:me_pf_a}: 
\beq\label{eq:bar_state_pf}
  \ket{\overline\Phi_F} \equiv e^{\hat \Omega_I(s)}e^{-\hat \Omega_F(s)} \ket{\Phi_F} = e^{\hat \Omega_I(s)} \ket{\Psi_F}\,.
\eeq

Using these newly defined states, we set up two schemes for evaluating the
transition matrix element: 
\beqn
 \text{PI:}\quad  M^{0\nu} &=& \bra{ \overline{\Phi}_F} e^{\hat \Omega_I } \hat O^{0\nu}e^{-\hat \Omega_I }\ket{ \Phi_I}\,, \label{eq:PI}\\
 \text{PF:}\quad  M^{0\nu} &=& \bra{ \Phi_F } e^{\hat \Omega_F } \hat O^{0\nu}e^{-\hat \Omega_F }\ket{\overline{\Phi}_I}\,. \label{eq:PF}
\eeqn
More explicitly, the procedures are as follows.  We begin with a GCM calculation
for the ground state of either the initial nucleus (in procedure PI) or the
final nucleus (in procedure PF) to obtain a reference state, and solve the flow
equation to obtain the corresponding unitary transformation operator $e^{\hat
\Omega_I}$ or $e^{\hat \Omega_F}$.  We then use the unitary transformation to
generate the evolved Hamiltonian $\hat H_{I/F}(s)$ and decay operator $\hat
O^{0\nu}_{I/F}(s)$.  Finally, we diagonalize the evolved Hamiltonian,
approximately, in the other nucleus --- the final nucleus in PI and the initial
nucleus in PF --- to obtain the barred state $\ket{\overline{\Phi}_F}$ or
$\ket{\overline{\Phi}_I}$.  This second diagonalization --- another GCM
calculation in our case --- would provide an exact result if it and the flow
were carried out without approximation.  Since the initial and final states are
(approximate) eigenvectors of the same Hamiltonian, we can
simply sandwich the corresponding evolved $0\nu\beta\beta$ operator between
those states, as in Eqs.\ \eqref{eq:PI} or \eqref{eq:PF}, to compute $M^{0\nu}$.
If we want, we can also use the evolved Hamiltonian to recompute the ground
state of the first nucleus, the one for which we solved the flow equations.  We
will show shortly that both the energies of low-lying states and the matrix
elements $M^{0\nu}$ can be improved in this way.

In either of the procedures above, one must use the BCH expansion \eqref{BCH} to
transform the charge-changing operator \eqref{eq:trans_op}.  In the present
work, we apply the NO2B approximation to each operator appearing in the BCH
series, \emph{including} general nested commutators $\left[\hat \Omega,\hat
O^{0\nu}\right]^{(n)}$, in the spirit of Ref.\ \cite{Morris15}. Dropping the
flow-parameter dependence for brevity, we see that the first commutator in the
series reads
\beqn
\label{CommutatorDBD}
[\hat\Omega, \hat O]
&=&[\hat\Omega^{(1)}, \hat O] + [\hat\Omega^{(2)}, \hat O]\\
&\equiv& \dfrac{1}{4}\sum_{pp'nn'} \left( O^{pp'}_{nn'}(\text{1B})+O^{pp'}_{nn'}(\text{2B})
\right) \nord{A^{pp'}_{nn'}} \,,
\eeqn
where the contributions involving the one-body and two-body parts of $\Omega$
are given by
\beqn
\label{eq:1bpart}
O^{pp'}_{nn'}(\text{1B})
&=& \sum_{p_1}\left[\Omega^{p}_{p_1} O^{p_1p'}_{nn'}+\Omega^{p'}_{p_1} O^{pp_1}_{nn'}\right] \nonumber\\
&&-\sum_{n_1}\left[\Omega^{n_1}_{n} O^{pp'}_{n_1n'}+\Omega^{n_1}_{n'} O^{pp'}_{nn_1}\right]
\,,
\eeqn
and 
\beqn\label{eq:trans_op_2B}
&O^{pp'}_{nn'}&(\text{2B})\nonumber\\
 &=&  \dfrac{1}{2}\sum_{p_1p_2} \Omega^{pp'}_{p_1p_2}   O^{p_1p_2}_{nn'}  (1-n_{p_1}-n_{p_2})\nonumber\\
&&
-\dfrac{1}{2}\sum_{n_1n_2} O^{pp'}_{n_1n_2} \Omega^{n_1n_2}_{nn'}
(1-n_{n_1}-n_{n_2}) \,,\nonumber\\
&\hphantom{=}&+
  \sum_{p_1n_1} (n_{p_1}-n_{n_1})  \left[
    \Omega^{n_1p'}_{n'p_1}  O^{p_1p}_{n_1n}
   -\Omega^{n_1p}_{n'p_1}   O^{p_1p'}_{n_1n}\right.
  \nonumber\\
&&\left.\qquad
+ \Omega^{n_1p}_{np_1}   O^{p_1p'}_{n_1n'}
- \Omega^{n_1p'}_{np_1}   O^{p_1p}_{n_1n'}  \right]\,,
\eeqn 
(cf.~Refs.~\cite{Hergert16,Hergert17}). Since $\hat\Omega(s)$ conserves charge,
no zero- or one-body terms are generated when we evaluate the commutator
\eqref{CommutatorDBD} (induced higher-body operators are truncated), and the
resulting operator has the same isospin structure as the initial transition
operator itself.  This means that we can use
Eqs.~\eqref{CommutatorDBD}--\eqref{eq:trans_op_2B} to recursively evaluate the
BCH series by replacing $\hat{O}$ with the appropriate nested commutator
$\left[\hat{\Omega},\hat{O}\right]^{(n)}$. Correlations in the reference state
only enter through fractional values of the occupation numbers, $0\leq n \leq
1$. At the currently employed NO2B truncation level, irreducible two- and
higher-body density matrices do not appear.

As discussed in Sec.~\ref{sec:ref}, our GCM reference states are projected onto
states with good angular momentum, allowing us to efficiently solve our
equations by working in a $J$-coupled scheme. Detailed expressions can be found
in Appendix \ref{append0}.

\section{Results and discussion}
\label{Results}

\begin{figure}[t]
\centering
\includegraphics[width=6cm]{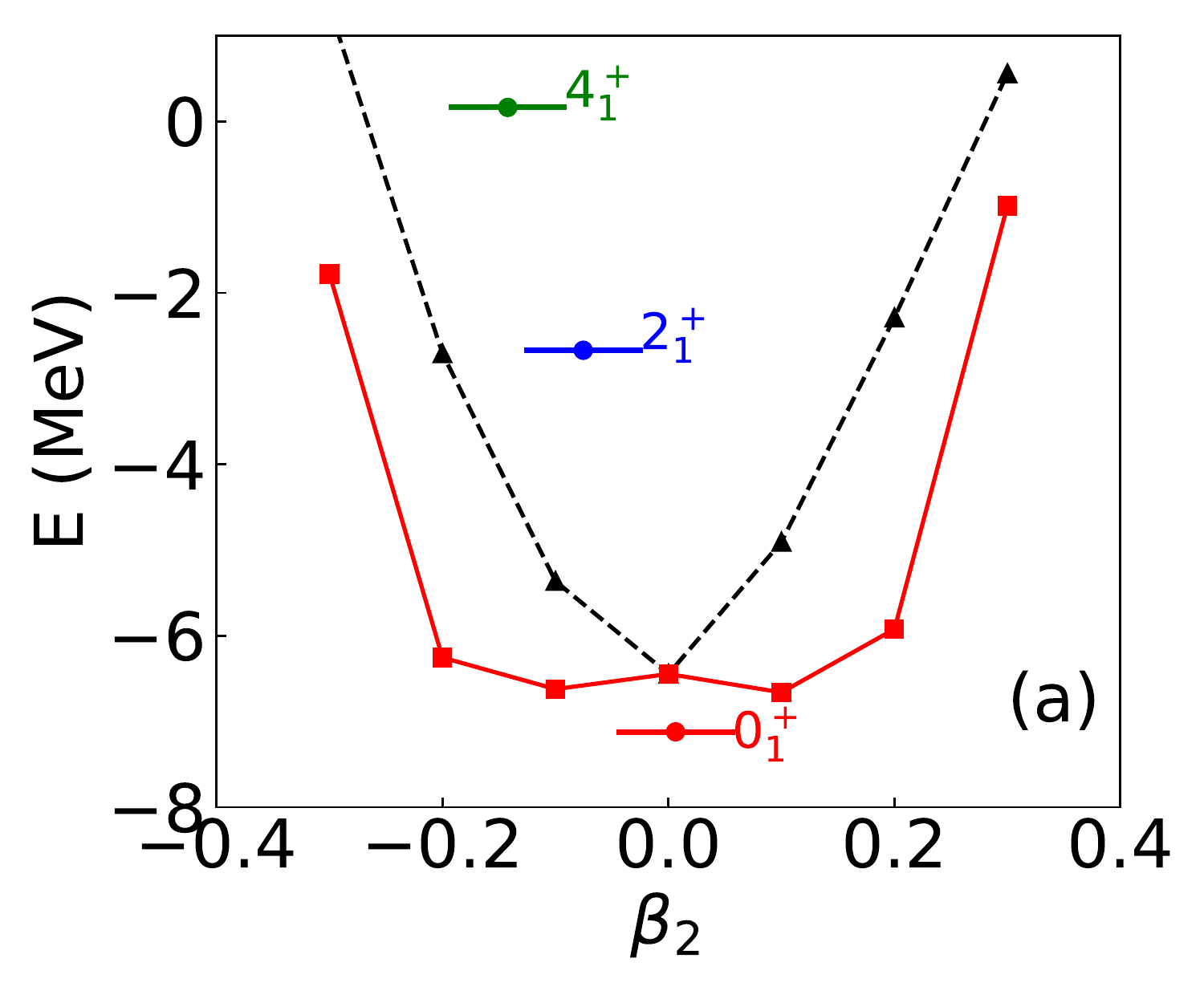}
\includegraphics[width=6cm]{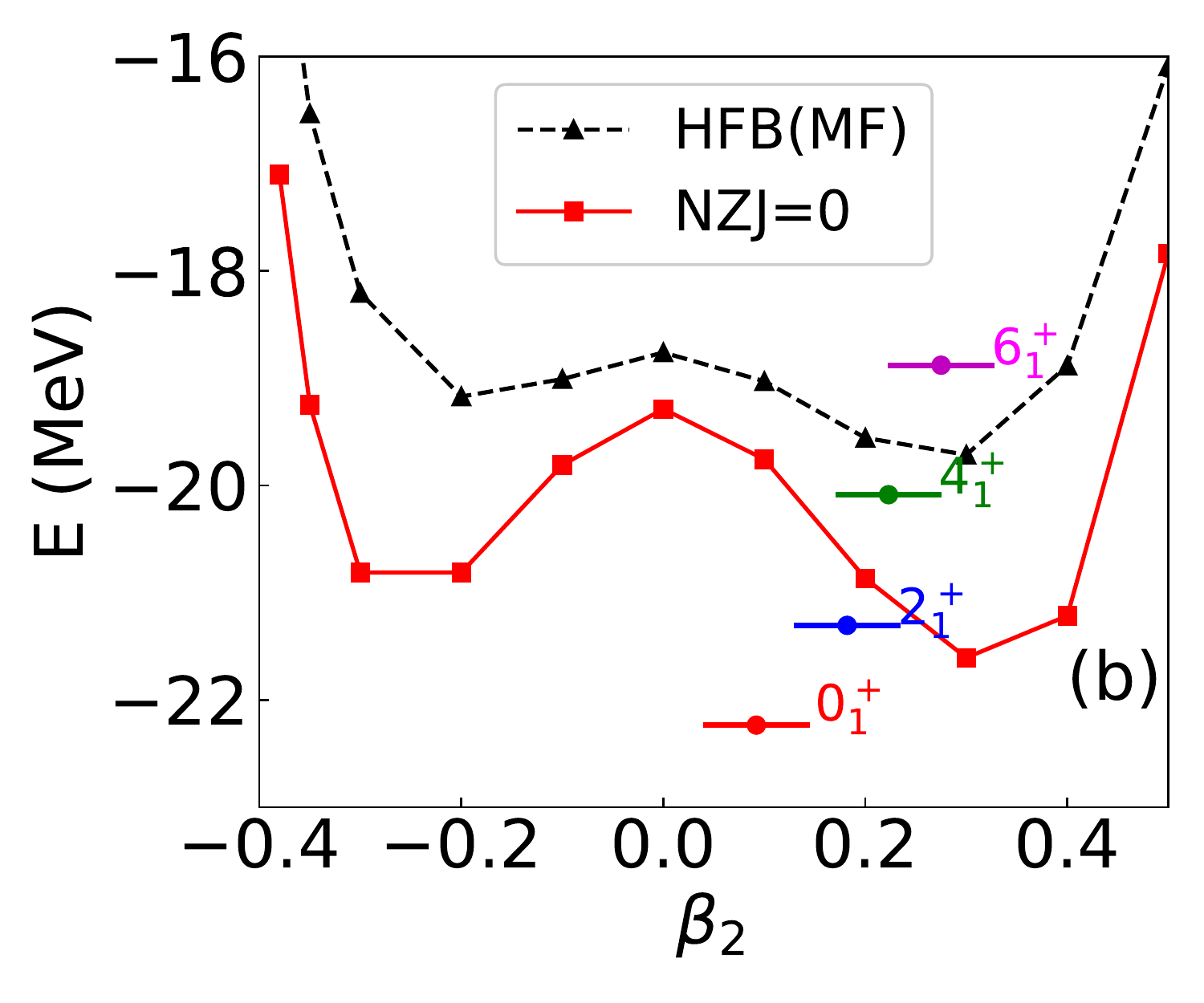}
\caption{(Color online) The energy surface from the projected HFB calculation
and the energies of low-lying states in $^{48}$Ca (a) and $^{48}$Ti (b). The
$x$-axis is the quadrupole deformation $\beta_2$. Each low-lying state from the
GCM calculation is placed at the average $\beta_2$ value for that state.}
\label{PES}
\end{figure}

\subsection{Energies of low-lying states}

Let us now apply the formalism described above to $^{48}$Ca and $^{48}$Ti,
within just the $fp$ shell (comprising the $0f_{7/2}, 0f_{5/2}, 1p_{3/2}$, and
$1p_{1/2}$ orbits) and with the interaction KB3G \cite{Poves01}. We aim to make
our GCM reference states as simple as possible while at the same time including
the most important collective correlations.  We therefore construct them from a
set of axially-deformed, angular-momentum- and particle-number-projected HFB
states with different values for the quadrupole deformation parameter $\beta_2
\equiv \chi\bra{\bq} (r/b)^2 Y_{20}\ket{\bq}/(\hbar\omega_0)$, with
$\hbar\omega_0=41.2 A^{-1/3}$ MeV, and $\chi=0.6$.  We let $\beta_2\in \{-0.3,
-0.2, \dots 0.2,0.3\}$ in $^{48}$Ca, and $\beta_2\in \{-0.3, -0.2 \dots 0.4,
0.5\}$ in $^{48}$Ti.  For these axially-deformed HFB states, one-dimensional
angular-momentum projection, together with particle-number projection, is
sufficient to restore all the broken symmetries.

\begin{figure}[t]
\centering
\includegraphics[width=4.2cm]{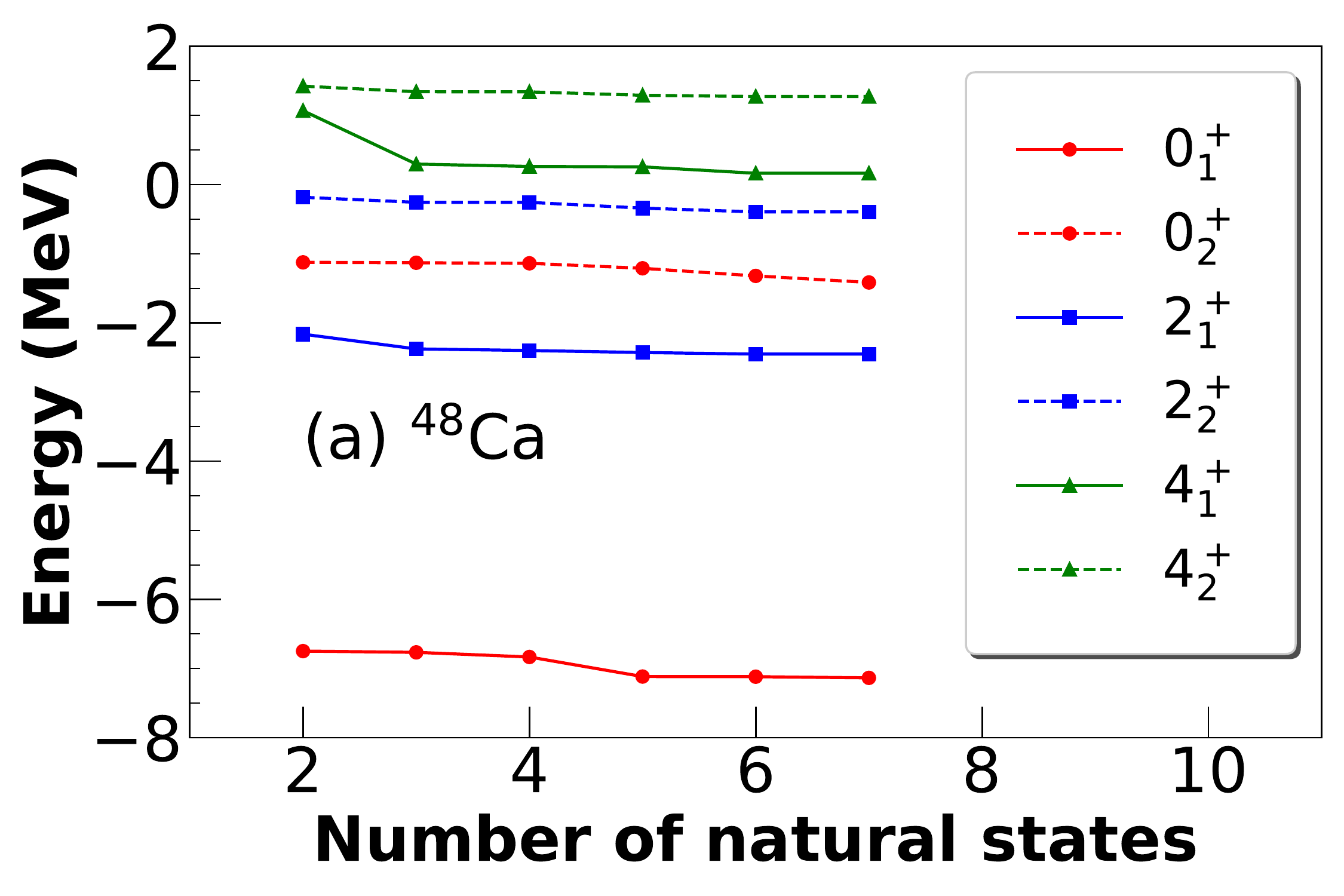}
\includegraphics[width=4.2cm]{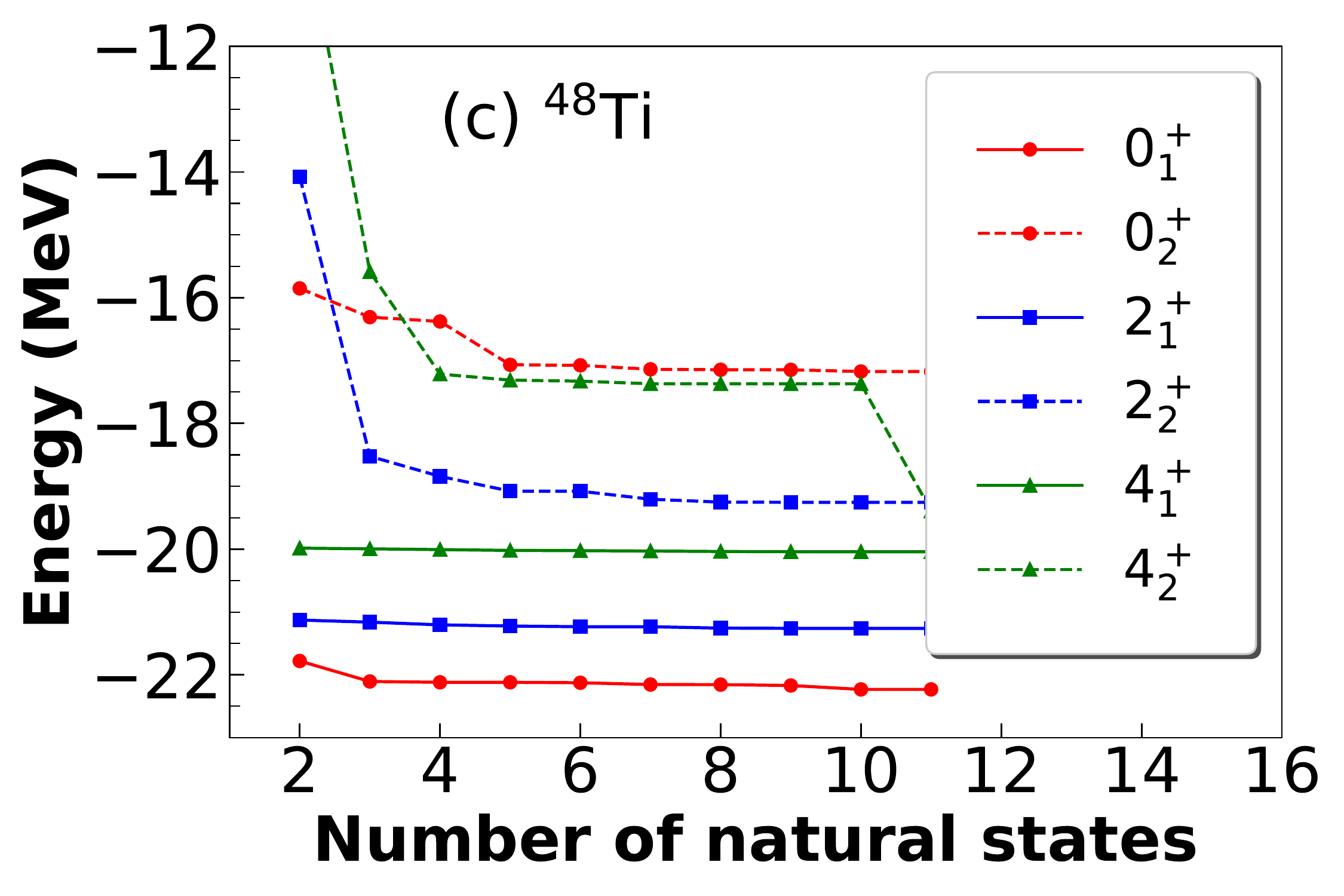}
\includegraphics[width=4.2cm]{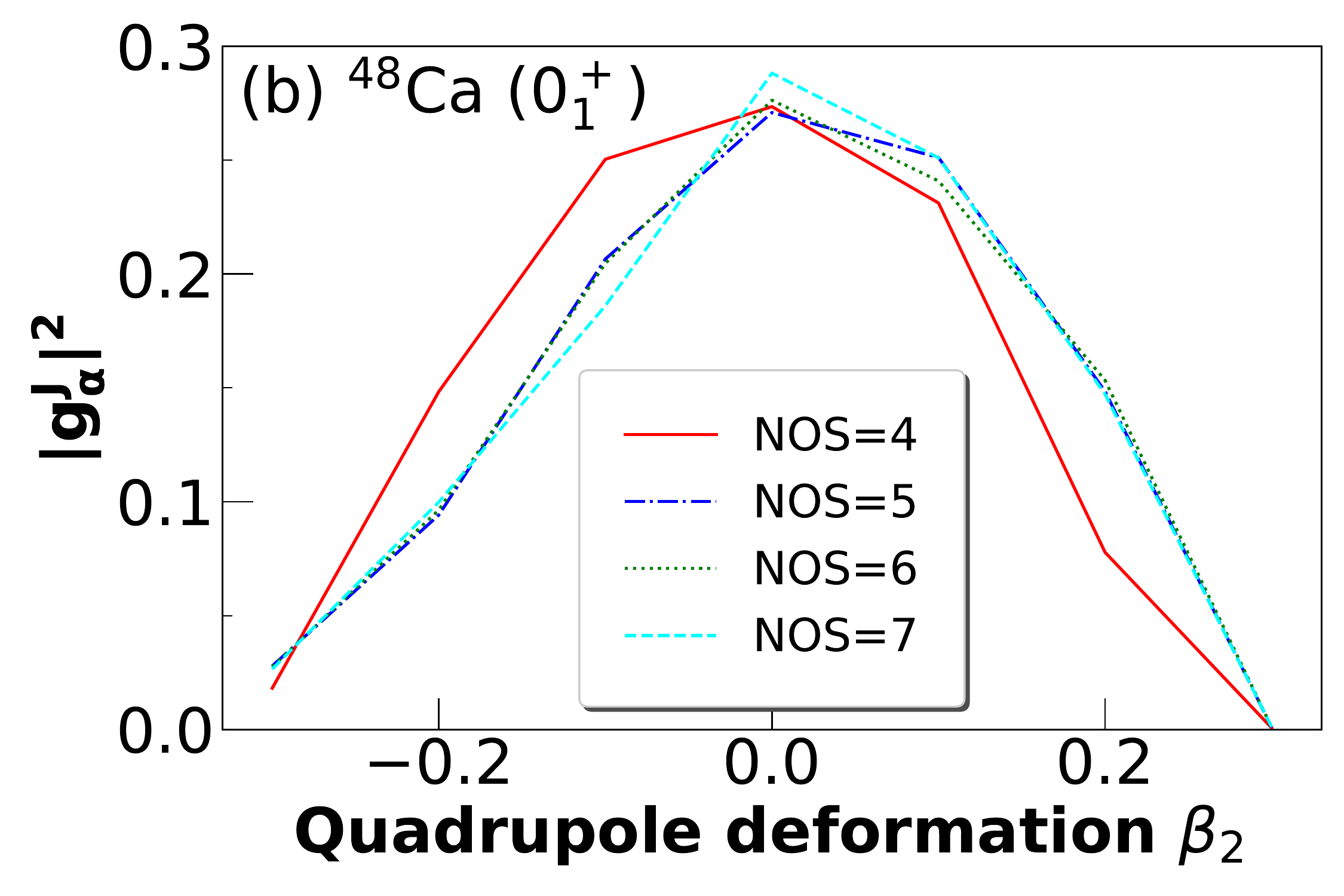}
\includegraphics[width=4.2cm]{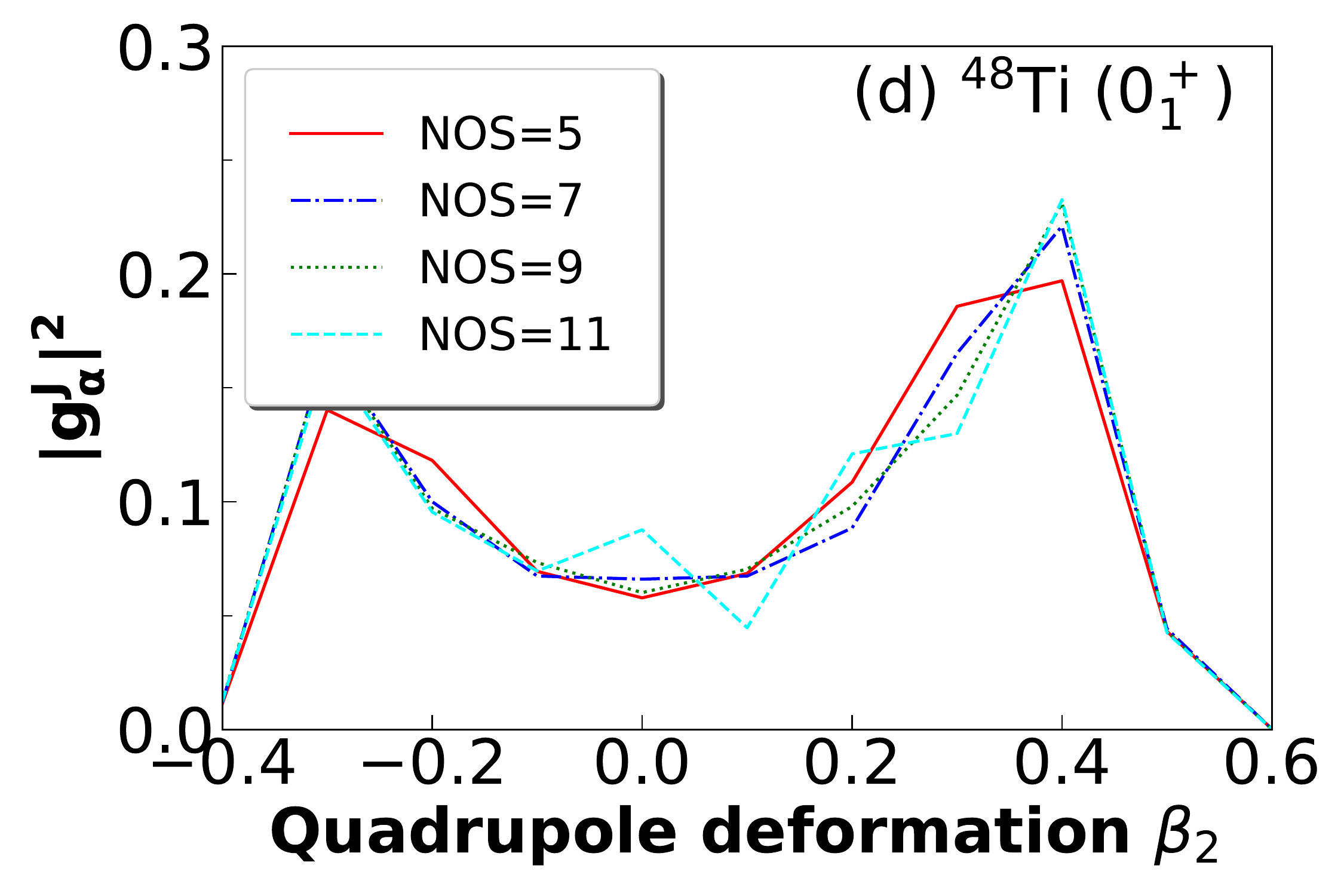}
\caption{(Color online) The energies of low-lying states in $^{48}$Ca (a) and
$^{48}$Ti (c) as a function of the number of natural states (NOS) adopted in the
GCM calculations.  The collective wave functions, defined as $g^{J}_\alpha
(\beta_2) = \sum_{\beta^\prime_2}
\left[\mathscr{N}^J\right]^{1/2}_{\beta_2,\beta^\prime_2}f^J_\alpha
(\beta^\prime_2)$, are shown for different choices of the NOS for the ground
state of $^{48}$Ca (b) and $^{48}$Ti (d), as a function of the quadrupole
deformation $\beta_2$.}
\label{convergence}
\end{figure}

Figure \ref{PES} presents curves of HFB energy vs,\ deformation (often referred
to as ``energy surfaces'' even in one dimension) for $^{48}$Ca and $^{48}$Ti,
both before and after projection onto states with $J=0$ and well-defined
particle number.  The global energy minimum is at a spherical shape in $^{48}$Ca
and a prolate shape in $^{48}$Ti. The figure also shows the energies of the
lowest lying states after the full calculations, which mix the shapes indicated
by the dots.  The ground states have GCM energies of $-7.12$ MeV in $^{48}$Ca
and $-22.18$ MeV in $^{48}$Ti. The results of exact diagonalization are $-7.57$
MeV and $-23.81$ MeV, both significantly smaller than the corresponding GCM
results.  Figure~\ref{convergence} shows that the energies of the low-lying
states are fairly stable against different choice of the number of natural
states (NOS) in the GCM calculations.  In other words, there are good
``plateaus" for the energies of both nuclei. The collective wave function for
the ground state is, however, somewhat sensitive to the NOS.

\begin{figure}[t]
\centering
\includegraphics[width=9cm]{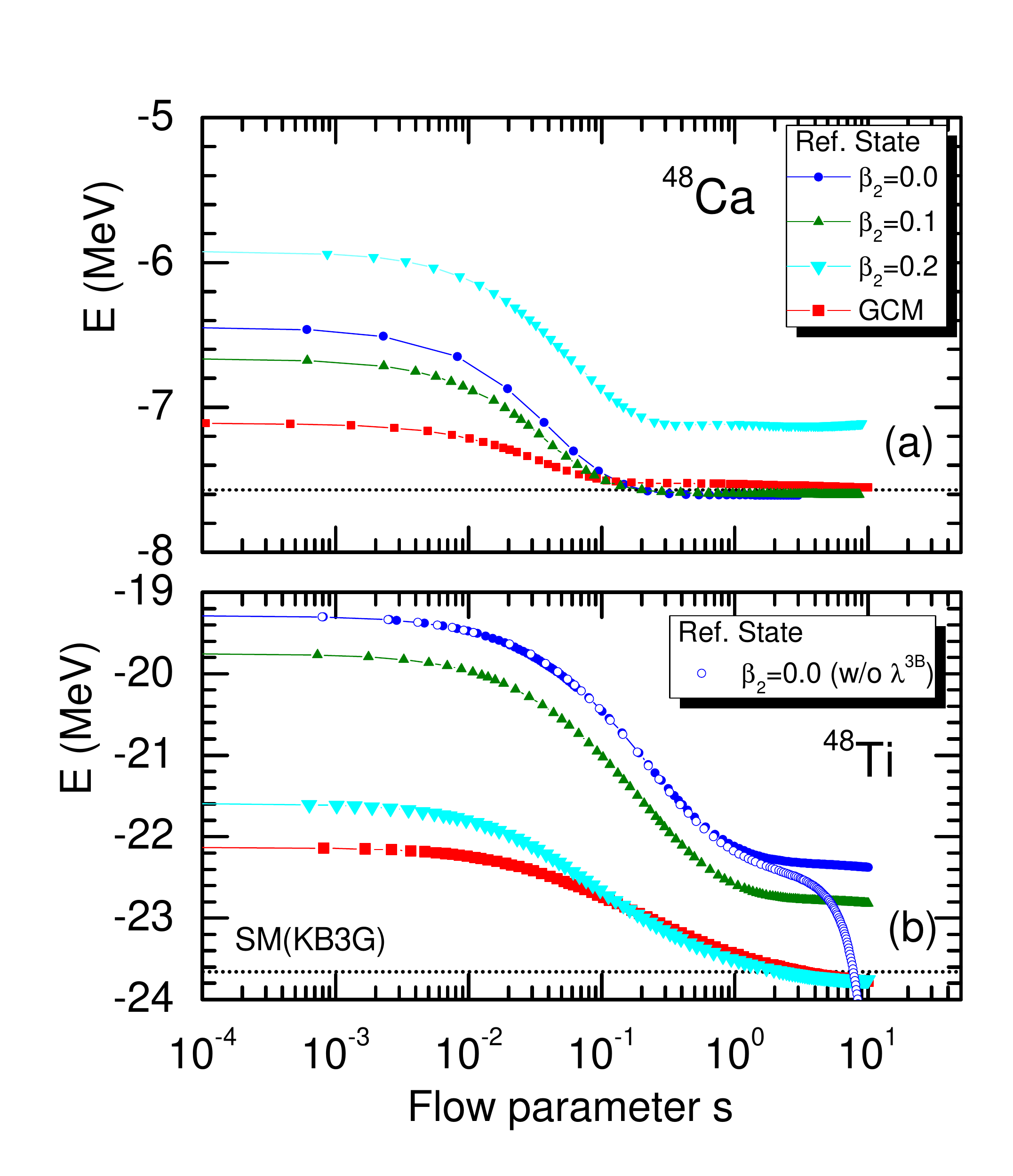}
\caption{(Color online) The ground-state energy $E$ a function of flow
parameter $s$ for $^{48}$Ca (a) and $^{48}$Ti (b), starting from either the
spherical reference state, a symmetry-projected HFB state, or a GCM state. The
horizontal line represents the energy from exact shell-model diagonalization.}
\label{flow}
\end{figure}

\begin{figure}[b]
\centering
\includegraphics[width=\columnwidth]{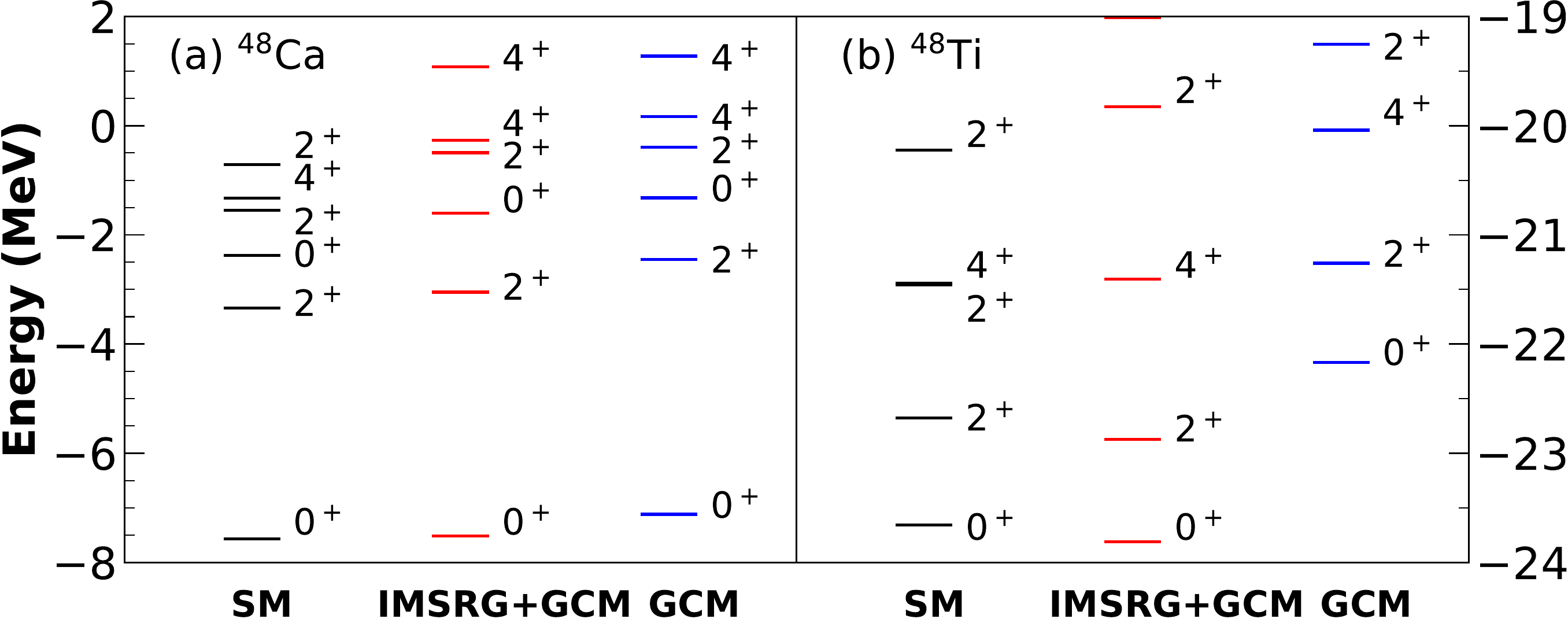}
\caption{(Color online) The low-lying states from the IMSRG+GCM and GCM alone,
with the KB3G interaction, in $^{48}$Ca (a) and $^{48}$Ti (b).  The exact
shell-model results \cite{Isacker17} are on the left in both panels.}
\label{Spectroscopy}
\end{figure}

Next we solve the IMSRG flow equations, starting both from these GCM states and
several simpler states, so that we can check the dependence of the results on
the reference.  Figure \ref{flow} shows the ground-state energy of $^{48}$Ca and
$^{48}$Ti, as a function of the flow parameter, starting from either the
spherical projected-HFB state, deformed projected-HFB states with $\beta_2=0.1,
0.2, (0.3)$, or the full GCM ground state $0^+_1$.  In $^{48}$Ca, except when
the reference state has $\beta_2=0.2$, the energy converges to almost the same
value, quite close to the result of exact diagonalization.  We note in passing
that, as discussed in Refs.~\cite{Hergert16,Hergert17,Hergert18} the IMSRG flow
may lead to an excited $0^+$ state that has a larger overlap with the reference
state than the ground state.  Because the energy of the reference state with
$\beta=0.2$ is lower than that of the $0^+_2$ state, the IMSRG, which cannot
raise the energy, does not converge to any sate at all.  But we have checked
that when we start from a reference state with $\beta_2=0.3$, the flow indeed
causes the energy to converge to that of the $0^+_2$ state.

In $^{48}$Ti, only projected-HFB reference states with $\beta \geq 0.2$ (and the
GCM state) lead to a final energy that is very close to the exact ground-state
value.  Starting from smaller values of $\beta$, we fall short of the correct
binding energy.  Clearly it is important that the reference state be deformed in
the right way; the IMSRG(2) flow by itself is not able to capture collective
correlations.  The bottom panel also shows that it is important to include
three-body irreducible densities in the flow equations.  If these are omitted,
as the pathological blue open symbols indicate, the energy fails to converge to
any value.

The final step in computing low-lying spectra, as we noted earlier, is to use
the evolved Hamiltonian from the IMSRG to carry out a second GCM calculation.
Figure \ref{Spectroscopy} compares the low-lying spectra from an initial GCM
calculation, from the second one (labeled IMSRG+GCM), and from exact
diagonalization.  The IMSRG+GCM energies are systematically lower than those
produced by the GCM alone, and are closer to the shell-model results
(mostly due to an overall shift).  The GCM is capable in principle of
reproducing the exact results with a sufficiently high number of
coordinates/basis states, but computation time scales badly with the number of
coordinates.  A more limited GCM calculation, followed by IMSRG evolution and a
second limited GCM calculation is much more efficient.  

Table \ref{Energy}, finally, contains the ground-state energies for $^{48}$Ca
and $^{48}$Ti in several approximation schemes.  The IMSRG+GCM overestimates the
energy of $^{48}$Ti by about $1\%$. This discrepancy is consistent with other
applications of the IMSRG in the NO2B approximation~\cite{Hergert17}, and
preliminary results suggest that it can be reduced significantly by using an
improved truncation scheme that accounts for induced three-body terms
\cite{Hergert18}.

\begin{table}
\tabcolsep=5pt
\caption{Ground-state energies (in MeV) for $^{48}$Ca and $^{48}$Ti, from
several calculations.}
\begin{tabular}{ccccc}
\hline
\hline
\\[-6pt]
             & SM     &  IMSRG+GCM&        GCM               & HFB(Sph.)\\[3pt]
\hline
\\[-8pt]
  $^{48}$Ca  & -7.57  & -7.56     &         -7.12            &  -6.45 \\
  $^{48}$Ti  & -23.66 & -23.81    &        -22.18            &  -18.76  \\[3pt]
\hline
\hline
\end{tabular}
\label{Energy}
\end{table}

\subsection{Matrix elements for neutrinoless double beta decay}
 
\begin{figure}[]
\centering
\includegraphics[width=8.5cm]{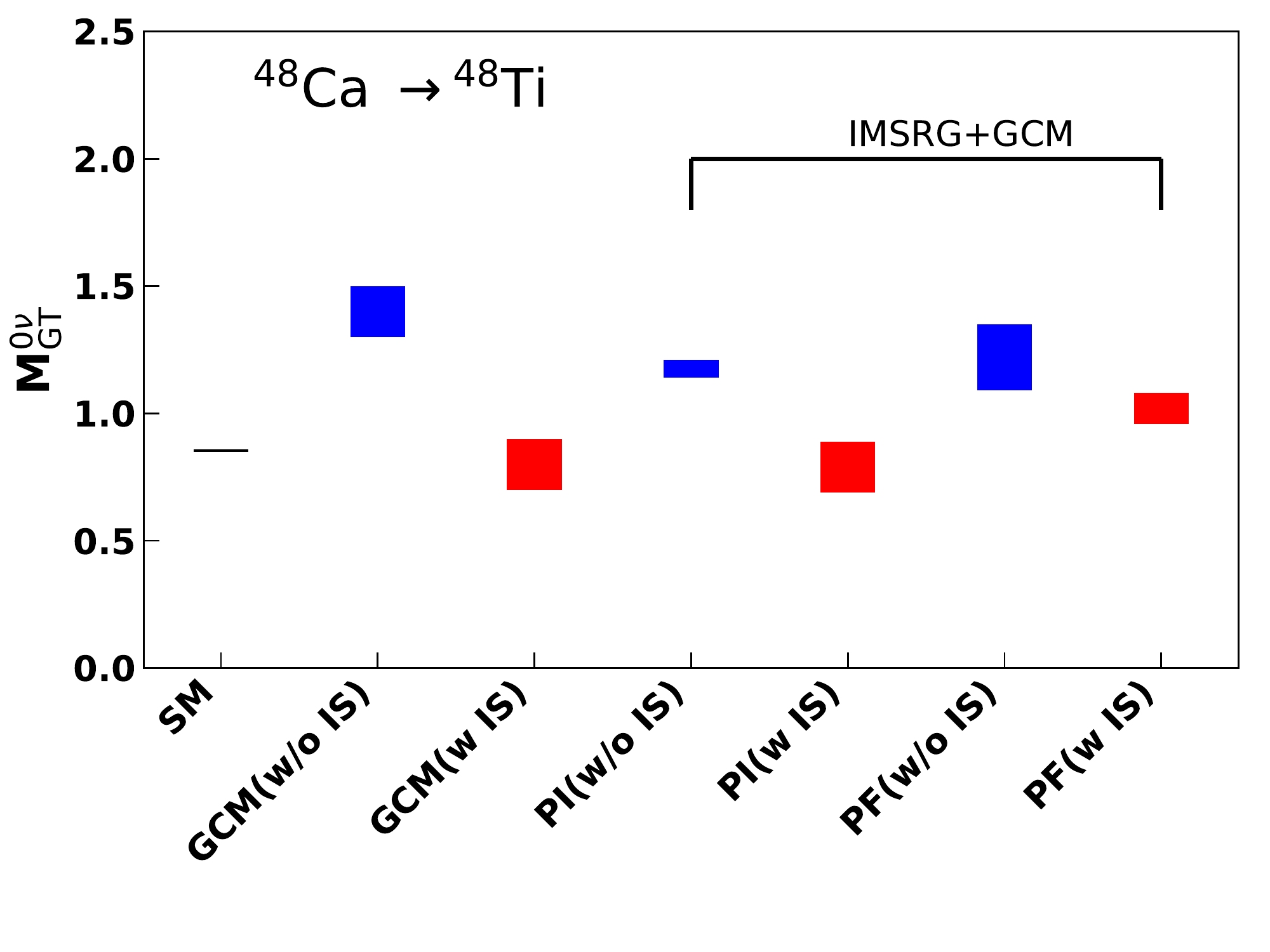}
\caption{(Color online) The Gamow-Teller part of $M^{0\nu}$ from several
calculations.  The blue boxes (w/o IS) are results of the GCM and IMSRG+GCM calculations
without an explicit isoscalar pairing coordinate, and the red boxes (w IS) are results
with that explicit coordinate.  
The uncertainty comes from the different choice  of natural states in the GCM calculation.}
\label{fig:NLDBD-GT}
\end{figure}

Figure\ \ref{fig:NLDBD-GT} compares the exact shell model result for the Gamow
Teller (GT) part of $M^{0\nu}$ to GCM and IMSRG+GCM results.  The blue boxes 
represent the results of the PI and PF procedures described above; the vertical
extent of the boxes represents the uncertainty in the optimal NOS, i.e.\ the
point at which to truncate the GCM basis before the energy becomes numerically
unstable.   In the previous GCM studies this uncertainty must also have existed, but was not explicitly investigated.
We show here that the matrix elements depend more on the NOS than does the energy.  The
dependence reflects the similar dependence of the collective wave functions
shown in Fig.\ \ref{convergence} .

The matrix elements produced by the two IMSRG+GCM procedures are in reasonable
agreement with one another, and both are slightly closer to the exact result
than the value produced by the GCM. The inability of the IMSRG evolution to
reduce the matrix element more significantly suggests that it is unable to fully
capture isoscalar pairing correlations, which shrink the matrix element
noticeably \cite{Hinohara14,Jiao17,Menendez16}.  The red boxes show the result
of including an isoscalar pairing amplitude as a GCM coordinate, in the manner
suggested by Refs.\ \cite{Hinohara14} and \cite{Jiao17}.  Now the agreement with
the exact result is good even before the IMSRG evolution, which does not spoil
it either. We note that including the isoscalar pairing amplitude as a generator coordinate introduces more redundancy  in the basis. By choosing the number of natural states properly, the low-lying eigenstates are at energies that are systematically somewhat smaller than without the isoscalar pairing coordinate.
Here we have run into the limits of what we can test in a
shell-model space.  Within the $fp$ shell, collective correlations, which
include isoscalar pairing, almost completely determine the $0\nu\beta\beta$
matrix element \cite{Menendez16}.  Because the IMSRG does not easily capture
these correlations, it has little effect on the matrix element; the collective
physics must thus all be built into the GCM state, an unsurprising situation.
In an ab-initio calculation in many shells, however, the situation is different.
Non-collective correlations from higher energy, including the short-range
correlations usually inserted by hand in shell-model calculations, will affect
the IMSRG operator evolution.  We expect our procedure(s) for computing
$0\nu\beta\beta$ matrix elements to work well in these kinds of calculations,
even if we are not able to prove it in a single shell.

Although neither IMSRG prescription affects the matrix element
very much, PF works slightly less well than PI, a result that is consistent with
the IMSRG ground-state energies in the two nuclei. The discrepancy with the
exact calculation is larger in $^{48}$Ti than in $^{48}$Ca, suggesting that the
approximate evolution with respect to a complicated GCM state containing both
valence neutrons and protons omits some important three-body contributions to
the like particle interaction, which determines the $^{48}$Ca energy and wave
function.

\section{Conclusion}
\label{Summary}

We have presented a very general framework for applying the IMSRG in conjunction
with GCM reference states to compute energies of low-lying states and
$0\nu\beta\beta$ matrix elements in nuclei with strong collective correlations,
including deformation.  Our method involves first a GCM calculation to generate a correlated reference state, 
then an IMSRG calculation, based on that state, to transform all operators, 
and finally a second GCM calculation that employs those operators.  
This approach allows us to use a single transformation to treat the transitions between two potentially quite different nuclei. 

We have benchmarked our method against the results of exact shell-model
diagonalization for $^{48}$Ca and $^{48}$Ti.  The IMSRG improves the
GCM-alone energies significantly, and the $0\nu\beta\beta$ matrix element
slightly, though in the one-shell calculations performed here the GCM
correlations by themselves are sufficient (and necessary) to nearly reproduce
the exact shell-model matrix element  when the coordinates
include the isoscalar pairing amplitude.  We are in the process of applying the
IMSRG+GCM in ab-initio calculations of this and other decays.

\section*{Acknowledgements}

We are grateful to S. Bogner, T. Morris, N. Parzuchowski, and S. R. Stroberg for
fruitful discussions.  We thank T. Rodriguez for sharing his unpublished GCM
code.  We also thank the Institute for Nuclear Theory at the University of
Washington for its hospitality. 

This material is based on work supported in part by the Scientific Discovery
through Advanced Computing (SciDAC) program funded by the U.S. Department of
Energy, Office of Science, Office of Advanced Scientific Computing Research and
Office of Nuclear Physics, under Award Number DE-SC0008641 (NUCLEI SciDAC
Collaboration), the U.S. Department of Energy, Office of Science, Office of
Nuclear Physics under Award Numbers DE-SC0017887, DE-FG02-97ER41019,
DE-SC0004142, and DE-SC0015376 (DBD Topical Theory Collaboration), and by the
National Natural Science Foundation of China under Grant No. 11575148.

Computing resources were provided by the U.S. National Energy Research Scientific Computing 
Center (NERSC), a DOE Office of Science User Facility supported by the Office of Science of 
the U.S.~Department of Energy under Contract No. DE-AC02-05CH11231.

 \begin{appendix}

\section{$J$-coupled operator evolution}
\label{append0}

It is convenient to rewrite the two-body matrix elements in the $J$-scheme with the relation \cite{Suhohen}
\beq
\ket{ kl} =\sum_{JM} \braket{ j_km_k j_lm_l| JM} [N_{KL}(J)]^{-1}\ket{ (KL)JM}
\,,
\eeq
where the normalized $J$-coupled two-body wave function is defined as
\beq
\ket{ (KL)JM} = N_{KL}(J) [a^\dagger_K a^\dagger_L]_{JM}\ket{ 0} \,,
\eeq
and the normalization factor is give by
$N_{KL}(J)=\sqrt{1+\delta_{KL}(-1)^J}/(1+\delta_{KL})$.  Here the capital letter
$K$ stands for the quantum numbers $\{\tau_k, n_k,l_k,j_k\}$.  With the above
definition, normalized $J$-coupled non-zero two-body matrix elements are related
to those in $M$-scheme as follows: 
\beqn
O^{J}_{(KL)(34)}&=&\sum_{m_k m_lm_3m_4} \braket{ j_km_k j_lm_l| JM} 
\braket{ j_3m_3 j_4m_4 | JM} \nonumber\\
&&\times\dfrac{1}{\sqrt{(1+\delta_{KL})(1+\delta_{34})}}   O^{kl}_{34}.
\eeqn
The {\em unnormalized} versions of the same matrix elements are given by
\beqn
&&\bar O^J_{(KL)(34)} \nonumber\\
&=& \sqrt{(1+\delta_{KL})(1+\delta_{34})} O^J_{KL34} \\
&=& \sum_{m_k m_lm_3m_4} \braket{j_km_k j_lm_l| JM} \braket{j_3m_3 j_4m_4|
JM}  O^{kl}_{34}. \nonumber
\eeqn
One can show that the  unnormalized $J$-coupled two-body matrix
elements  corresponding to the first two terms in Eq.(\ref{eq:trans_op_2B}) ($pp$ parts)  are given by
\beqn
&&\bar {\cal O}^J_{(KL)(34)}(pp)\nonumber\\
&=& \dfrac{1}{2}\sum_{CD} \bar \Omega^J_{(KL)(CD)} \bar
O^J_{(CD)(34)} (1-n_c-n_d)\nonumber\\
&&-\dfrac{1}{2}\sum_{12}  \bar O^J_{(KL)(12)} \bar
\Omega^J_{(12)(34)} (1-n_1-n_2).
\eeqn 
and those corresponding to the last two terms in Eq.(\ref{eq:trans_op_2B}) ($ph$ parts) are
\begin{widetext}
\beqn
\bar {\cal O}^J_{(KL)(34)}(ph)
&=&
- \sum_{J'}\hat {J^\prime}^2
\left\{\begin{array}{ccc}
j_k & j_l & J \\
j_3 & j_4 & J'
\end{array}\right\} 
\sum_{A6}   (n_6-n_a)
\bar O^{J'}_{(K\bar 4) (6\bar A)}  
\bar \Omega^{J'}_{(6\bar A)(3\bar L)}     
\\
&&-(-1)^{j_k+j_l+J+1}\sum_{J'}\hat {J^\prime}^2
\left\{\begin{array}{ccc}
j_l & j_k & J \\
j_3 & j_4 & J'
\end{array}\right\}
\sum_{A6}(n_6-n_a)
\bar O^{J'}_{(L\bar 4) (6\bar A)} 
\bar \Omega^{J'}_{(6\bar A) (3\bar K)}\nonumber\\
&&
+\sum_{J'}\hat {J^\prime}^2  \left\{\begin{array}{ccc}
j_k & j_l & J \\
j_3 & j_4 & J'
\end{array}\right\}
\sum_{A6}
(n_a-n_6)
\bar \Omega^{J'}_{(K\bar 4)(A\bar 6)}
\bar O^{J'}_{(A\bar 6) (3\bar L)} \nonumber\\
&&+ (-1)^{j_k+j_l+J+1}  \sum_{J'}\hat {J^\prime}^2
\left\{\begin{array}{ccc}
j_l & j_k & J \\
j_3 & j_4 & J'
\end{array}\right\}
\sum_{A6}
(n_a-n_6)
\bar \Omega^{J'}_{(L\bar 4) (A\bar 6)}
\bar O^{J'}_{(A\bar 6) (3\bar K)}\nonumber
\eeqn
\end{widetext}
where  the Latin indices $k, l$ stand for proton states and the numerals 3,4 stand for neutron states.
Only the $\Omega$-matrix elements of the form $\Omega^{np}_{n'p'}$  contribute to the $ph$-parts of ${\cal O}$.  The unnormalized
$ph$ matrix element $\bar O^J$ is related to that of $pp$ matrix element $\bar
O^J$ by the Pandya transformation \cite{Suhohen}
 \beqn
&&   \bar O^{J}_{ (\alpha\bar \beta) (\gamma \bar \delta)}
=-\sum_{J^{'}}  {\hat J^{'2}}
\left\{\begin{array}{ccc}
j_\alpha & j_\beta & J \\
j_\gamma & j_\delta & J^{'}
\end{array}\right\}
\bar {\cal O}^{J'}_{(\alpha \delta) (\gamma \beta)}.
\eeqn

\section{The Brillouin generator}
\label{append1}
 
In the IMSRG(2) calculation, we truncate the matrix elements of $\eta(s)$ at the
NO2B level,
\beq
\hat \eta(s) = \sum_{ij}\eta^k_l(s)  \{ A^k_l \} +\dfrac{1}{4}
\sum_{klmn}\eta^{kl}_{mn}(s)\{ A^{kl}_{mn} \} \,, 
\eeq
and use the Brillouin generator \cite{Hergert17}. The matrix elements of the one and two-body parts are ($\bar n_i = 1- n_i$)

\beqn
\eta^k_l
&=&f^l_k (n_l-n_k ) -\dfrac{1}{2}\sum_{abc}(\Gamma^{la}_{bc}\lambda^{ka}_{bc}-\Gamma^{ab}_{kc}\lambda^{ab}_{lc}),\\
\eta^{kl}_{mn}
&=&\Gamma^{mn}_{kl}(\bar n_k\bar n_ln_mn_n-n_k n_l\bar n_m\bar n_n) \nonumber\\
&&
+\sum_a(f^a_k\lambda^{al}_{mn}+f^a_l\lambda^{ka}_{mn}-f^m_a\lambda^{kl}_{an}-f^n_a\lambda^{kl}_{ma})\nonumber\\
&&+\dfrac{1}{2}[(\lambda\Gamma)^{mn}_{kl}(1-n_k-n_l)-
(\Gamma\lambda)^{mn}_{kl}(1-n_m-n_n)] \nonumber\\
&&+(1-\hat P_{mn})(1-\hat P_{kl})\sum_{ac}
\Gamma^{am}_{cl}\lambda^{ak}_{cn}(n_l-n_m) \nonumber\\
&&+\dfrac{1}{2}\sum_{abc}[(1-\hat
P_{mn})\Gamma^{ma}_{bc}\lambda^{akl}_{bcn}+(1-\hat
P_{kl})\Gamma^{ab}_{lc}\lambda^{abk}_{cmn}].\nonumber\\ 
\eeqn
We use the $J$-coupled scheme above to save memory.  Since the terms involving the
three-body irreducible density are more complicated than the others, we write
them explicitly:
\begin{widetext}
\bsub
\beqn  
\bar\eta^J_{(KL)(MN)}(\lambda^{3B},1)   
&=& \dfrac{1}{2}  \sum_{ABC}\sum_{J_2J_{ak}J_{bc}}(-1)^{J_{ak}+j_a-j_k}  
 \hat J_{ak}\hat J_{bc}\hat J^2_2
 (-1)^{J_{ak}+J_{bc}+j_l+j_n}   \nonumber\\
 &&\times
  \left\{
 \begin{array}{ccc}
 j_a  & J_{ak} & j_{k} \\
 j_l & J        & J_2\\
 \end{array}
 \right\}
  \left\{
 \begin{array}{ccc}
 j_a  & J_{bc} & j_{m} \\
 j_n & J        & J_2\\
 \end{array}
 \right\} 
 \bar\Gamma^{J_{bc}}_{(MA)(BC)} 
\bra{ (j_aj_k) J_{ak} j_l; J_2} \lambda \ket{ (j_bj_c)J_{bc} j_n; J_2},  \\
 \bar\eta^J_{KL)(MN)}(\lambda^{3B},2)  
&=&(-1)^{j_m+j_n-J+1} \dfrac{1}{2} \sum_{ABC}
\sum_{J_{ak}J_{bc}}\sum_{J_2}(-1)^{j_a+J_{ak}-j_k} 
(-1)^{4j_a+J_{bc}+J_{ak}+2J+j_m+j_l} \hat J^2_2  \hat J_{bc}   \hat J_{ak}\nonumber\\
&&\times
\left\{\begin{array}{ccc}
j_a & J_{ak} & j_{k} \\
j_l & J & J_{2} \\
\end{array}
\right\}  \left\{\begin{array}{ccc}
j_a & J_{bc} & j_{n} \\
j_m & J & J_{2} \\
\end{array}
\right\}   \bar\Gamma^{J_{bc}}_{(NA)(BC)}   \bra{ (j_aj_k)J_{ak}j_l; J_2 }
\lambda\ket{ (j_bj_c)J_{bc}j_m;J_2 },\\ 
 \bar\eta^J_{(KL)(MN)}(\lambda^{3B},3)   
&=& \dfrac{1}{2}  \sum_{ABC}\sum_{J_2J_{ab}J_{cm}}(-1)^{j_c+J_{cm}-j_m} 
 \hat J_{ab}\hat J_{cm}\hat J^2_2 (-1)^{J_{ab}+J_{cm}+J+j_l+j_n+2j_k}
\nonumber\\
&& \times \left\{
 \begin{array}{ccc}
 j_c  & J_{cm} & j_{m} \\
 j_n & J        & J_2\\
 \end{array}
 \right\}
  \left\{
 \begin{array}{ccc}
 j_c  & J_{ab} & j_{l} \\
 j_k & J        & J_2\\
 \end{array}
 \right\}  
  \bar\Gamma^{J_{ab}}_{(AB)(LC)} 
 \bra{ (j_aj_b) J_{ab} j_k; J_2} \lambda \ket{ (j_cj_m)J_{cm} j_n; J_2},\\
 \bar\eta^J_{(KL)(MN)}(\lambda^{3B},4)  
&=& (-1)^{j_k+j_l-J+1}\dfrac{1}{2}
 \sum_{ABC} \sum_{J_{ab} J_{cm}J_2}(-1)^{j_c+J_{cm}-j_m} 
 (-1)^{2j_l+4j_c+j_k+j_n+J_{ab}+ J_{cm}+J} 
 \hat J_{cm}\hat J_{ab} \hat J^2_2 \nonumber\\
 &&\times\left\{\begin{array}{ccc}
 j_c & J_{ab} & j_{k} \\
 j_l & J & J_{2} \\
 \end{array}
 \right\}
 \left\{\begin{array}{ccc}
 j_c & J_{cm} & j_{m} \\
 j_n & J & J_{2} \\
 \end{array}
 \right\}  \bar\Gamma^{J_{ab}}_{(AB)(KC)} 
 \bra{ (j_aj_b)J_{ab}j_l; J_2 } \lambda\ket{ (j_cj_m)J_{cm}j_n;J_2 } \,. 
\eeqn
\esub
\end{widetext}
Here, $\lambda^{3B}$ in parentheses indicates a dependence on the irreducible three-body density $\bra{ (j_1j_2) J_{12} j_3; J}
\lambda \ket{ (j_4j_5)J_{45} j_6; J}$, the calculation of which is given in Appendix (\ref{append2}).

 \section{Density matrices of multi-reference states}
 \label{append2}

We present here the most important expressions needed to compute the density
matrices associated with a general multi-reference state, taken here to 
have spin and parity $0^+$.  The irreducible (or residual) one-, two-, and
three-body parts of density matrix elements follow from a cumulant expansion (\ref{cumulants}).
In the
coupled scheme, the expressions for the one- and two-body densities take the
form (with $\kappa=\{n,\alpha\}$)\,,
\beqn
\lambda^{J=0}_{\kappa_1\kappa_2}
&=&\rho^{J=0}_{\kappa_1\kappa_2}
\equiv \dfrac{[a^\dagger_{\kappa_1}\tilde
a_{\kappa_2}]^0_0}{\sqrt{2j_1+1}}\delta_{\alpha_1\alpha_2}\,.\\
\lambda^J_{(12)(34)} 
&=& \rho^J_{(12)(34)} - \lambda^{J=0}_{\kappa_1,\kappa_3} \lambda^{J=0}_{\kappa_2,\kappa_4}\delta_{\alpha_1,\alpha_3}\delta_{\alpha_2,\alpha_4}\nonumber\\
&& +(-1)^{J-(j_1+j_2)}  \lambda^{J=0}_{\kappa_1,\kappa_4} \lambda^{J=0}_{\kappa_2,\kappa_3}\delta_{\alpha_1,\alpha_4}\delta_{\alpha_2,\alpha_3} ,
\eeqn
where $\alpha=\{\tau lj\}$, and the expression for the irreducible three-body
density takes the form 

 \begin{widetext}
 \beqn
 &&\bra{ (j_1j_2)J_{12}j_3;J_{123}} \lambda\ket{  (j_4j_5)J_{45}j_6;J_{123}}\nonumber\\ 
 &=&\sum_{m_1,m_2,\cdots,m_6}
 \braket{ j_1m_1j_2m_2 | J_{12}M_{12}}
 \braket{ J_{12}M_{12} |j_3m_3 J M}  
 \braket{ j_4m_4j_5m_5| J_{45}M_{45}}
 \braket{ J_{45}M_{45} j_6m_6| J M} 
 \lambda^{123}_{456}  \nonumber\\
  &=& \bra{ (j_1j_2)J_{12}j_3;J_{123}} \rho\ket{  (j_4j_5)J_{45}j_6;J_{123}}
  -\sum^{15}_{i=1} T_i \,,
 \eeqn
 where
 \bsub
 \beqn
 T_1 &=& (-1)^{J_{45}+j_2+j_3+1} \hat J_{12}\hat J_{45}
  \left\{\begin{array}{ccc}
 j_1 & j_2 & J_{12} \\
 j_3 & J & J_{45} \\
 \end{array}
 \right\}\rho^{J=0}_{\kappa_1\kappa_6} \rho^{J_{45}}_{(23)(45)}\,,
  \eeqn
  \beqn
 T_2 &=& \sum_{J_{23}} (-1)^{J_{45}+J_{23}+j_1+j_2+j_3+j_4}
  \hat J_{12}\hat J_{45}\hat J^2_{23}
  \left\{\begin{array}{ccc}
 j_4 & j_1 & J_{45} \\
 J & j_6 & J_{23} \\
 \end{array}
 \right\} \left\{\begin{array}{ccc}
 j_2 & j_3 & J_{23} \\
 J & j_1 & J_{12} \\
 \end{array}
 \right\}\rho^{J=0}_{\kappa_1\kappa_5} \rho^{J_{23}}_{(23)(64)}\,,
 \eeqn
 \beqn
 T_3 =
  \sum_{J_{23}}(-1)^{j_2+j_3+j_5+j_6}
  \hat J_{12}\hat J_{45}\hat J^2_{23}
  \left\{\begin{array}{ccc}
 j_5 & j_1 & J_{45} \\
 J & j_6 & J_{23} \\
 \end{array}
 \right\} \left\{\begin{array}{ccc}
 j_2 & j_3 & J_{23} \\
 J & j_1 & J_{12} \\
 \end{array}
 \right\}\rho^{J=0}_{\kappa_1\kappa_4} \rho^{J_{23}}_{(23)(56)}\,,
 \eeqn
 \beqn
 T_4
 &=&(-1)^{j_1+j_2-J_{12}+1}
  \hat J_{12}\hat J_{45}
  \left\{\begin{array}{ccc}
 j_2 & j_1 & J_{12} \\
 j_3 & J & J_{45} \\
 \end{array}
 \right\} \rho^{J=0}_{\kappa_2\kappa_6} \rho^{J_{45}}_{(31)(45)}\,,
 \eeqn
 \beqn
 T_5 &=&
  \sum_{J_{31}}(-1)^{j_4+j_1+J_{12}+J_{45}+1}
   \hat J_{12}\hat J_{45}\hat J^2_{31}
  \left\{\begin{array}{ccc}
 j_4 & j_2 & J_{45} \\
 J & j_6 & J_{31} \\
 \end{array}
 \right\} \left\{\begin{array}{ccc}
 j_1 & j_3 & J_{31} \\
 J & j_2 & J_{12} \\
 \end{array}
 \right\}\rho^{J=0}_{\kappa_1\kappa_4} \rho^{J_{31}}_{(31)(64)}\,,
 \eeqn
 \beqn
 T_6 &=&
 \sum_{J_{31}}(-1)^{j_1+j_2+J_{12}+j_5+j_6+J_{31}}
   \hat J_{12}\hat J_{45}\hat J^2_{31}
  \left\{\begin{array}{ccc}
 j_5 & j_2 & J_{45} \\
 J & j_6 & J_{31} \\
 \end{array}
 \right\} \left\{\begin{array}{ccc}
 j_1 & j_3 & J_{31} \\
 J & j_2 & J_{12} \\
 \end{array}
 \right\}\rho^{J=0}_{\kappa_2\kappa_4} \rho^{J_{31}}_{(31)(56)}.
 \eeqn
 \beqn
 T_7
 &=&  \delta_{J_{12}J_{45}}
   \rho^{J=0}_{\kappa_3\kappa_6} \rho^{J_{12}}_{(12)(45)}\,,
 \eeqn
 \beqn
 T_8
 &=&(-1)^{j_3+j_4+J_{45}+1}
   \hat J_{12}\hat J_{45}
  \left\{\begin{array}{ccc}
 j_4 & j_3 & J_{45} \\
 J & j_6 & J_{12} \\
 \end{array}
 \right\} \rho^{J=0}_{\kappa_3\kappa_5} \rho^{J_{12}}_{(12)(64)}\,,
 \eeqn
  \beqn
  T_9 &=& (-1)^{J_{12}+j_5+j_6+1} \hat J_{12}\hat J_{45}
  \left\{\begin{array}{ccc}
 j_5 & j_3 & J_{45} \\
 J & j_6 & J_{12} \\
 \end{array}
 \right\} \rho^{J=0}_{\kappa_3\kappa_4} \rho^{J_{12}}_{(12)(56)}\,,
  \eeqn
 \beqn
 T_{10}
 &=&-2\hat J_{12}\hat J_{45}
  \left\{\begin{array}{ccc}
 j_1 & j_2 & J_{12} \\
 j_3 & J & J_{45} \\
 \end{array}
 \right\} \rho^{J=0}_{\kappa_1\kappa_6}
 \rho^{J=0}_{\kappa_2\kappa_5}\rho^{J=0}_{\kappa_3\kappa_4}\,,
 \eeqn
 \beqn
 T_{11}
 &=&-2(-1)^{J_{45}+j_2+j_3+1}\hat J_{12}\hat J_{45}
  \left\{\begin{array}{ccc}
 j_1 & j_2 & J_{12} \\
 j_3 & J & J_{45} \\
 \end{array}
 \right\}
 \rho^{J=0}_{\kappa_1\kappa_6}
 \rho^{J=0}_{\kappa_2\kappa_4}\rho^{J=0}_{\kappa_3\kappa_5}\,,
 \eeqn
 \beqn
 T_{12}
 &=&-2(-1)^{j_1+j_2-J_{12}+1}\hat J_{12}\hat J_{45}
  \left\{\begin{array}{ccc}
 j_2 & j_1 & J_{12} \\
 j_3 & J & J_{45} \\
 \end{array}
 \right\}
 \rho^{J=0}_{\kappa_1\kappa_5}
 \rho^{J=0}_{\kappa_2\kappa_6}\rho^{J=0}_{\kappa_3\kappa_4}\,,
 \eeqn
 \beqn
 T_{13}&=&
  2(-1)^{j_1+j_2-J_{12}}  \delta_{J_{12}J_{45}}
 \rho^{J=0}_{\kappa_1\kappa_5}
 \rho^{J=0}_{\kappa_2\kappa_4}\rho^{J=0}_{\kappa_3\kappa_6}\,,
 \eeqn
 \beqn
 T_{14}
 &=&2(-1)^{j_2+j_3-J_{12}+J_{45}}\hat J_{12}\hat J_{45}
  \left\{\begin{array}{ccc}
 j_2 & j_1 & J_{12} \\
 j_3 & J & J_{45} \\
 \end{array}
 \right\}
 \rho^{J=0}_{\kappa_1\kappa_4}
 \rho^{J=0}_{\kappa_2\kappa_6}\rho^{J=0}_{\kappa_3\kappa_5}\,,
 \eeqn
 \beqn
 T_{15}
 &=&
  -2 \delta_{J_{12}J_{45}}\rho^{J=0}_{\kappa_1\kappa_4}
  \rho^{J=0}_{\kappa_2\kappa_5}\rho^{J=0}_{\kappa_3\kappa_6}\,.
 \eeqn
 \esub
 \end{widetext}

\end{appendix}

\end{document}